\documentclass[aps,pra,preprint,groupedaddress]{revtex4} 

\usepackage{graphicx}
\usepackage{bm}
\usepackage{array}
\usepackage{amsmath}

\begin{document}


\title{Numerical modeling of collisional dynamics of Sr in an optical dipole trap}



\author{M. Yan, R. Chakraborty, A. Mazurenko, P. G. Mickelson, Y. N. Martinez de Escobar, B. J. DeSalvo, and T. C. Killian}

\affiliation{Rice University, Department of Physics and Astronomy, Houston,
Texas, 77251}




\date{\today}

\begin{abstract}
We describe a model of inelastic and elastic collisional dynamics of atoms in an
optical dipole trap that utilizes numerical evaluation of statistical
mechanical quantities and numerical solution of equations for the evolution of
number and temperature of trapped atoms. It can be used for traps that possess
little spatial symmetry and when the ratio of trap depth to sample temperature
is relatively small. We
compare simulation results with experiments
on $^{88}$Sr and $^{84}$Sr, which have well-characterized collisional properties.

\end{abstract}


\maketitle



\section{Introduction\label{section:introduction}}




Understanding the collisional dynamics of trapped, ultracold atoms is essential
for optimizing forced evaporative cooling \cite{hes86,tom86} and obtaining
quantum degenerate Bose \cite{dgp99} and Fermi gases \cite{gps08}. It also
allows determination of ultracold collision properties from the evolution of
number and temperature  in a trapped sample of atoms or molecules
\cite{bgm97,jtl06,kre05}.

Many recipes have been presented for relating the evolution of the trapped gas
to underlying physical parameters. Typically the collisional dynamics are
described by differential equations for the time rate of change of the atom
number ($N$) and  and total energy ($E$), as originally suggested by \cite{hes86,tom86}.
The method has been extended and developed in many other works
\cite{dmk95,kvd96,lrw96,ber97,thv04,hdj00,sbh97,ogg01}.
The standard treatment of evaporation is described by Luiten \textit{et al.}
\cite{lrw96}, which derives expressions for thermodynamic quantities from the
kinetic equations using an assumption of sufficient ergodicity and a truncated
Boltzmann velocity distribution. Analytic evaluation of these expressions is
straightforward for power-law traps. Noteworthy subsequent improvements over
this work include the addition of effects of time-dependent potentials
\cite{ber97}, energy-dependent cross sections \cite{thv04}, and quantum
statistics \cite{hdj00}. Prescriptions have  been offered for optimizing
evaporation \cite{sbh97} and deriving scaling laws \cite{ogg01}. Direct Monte
Carlo simulations have also been presented to relax the assumption of
sufficient ergodicity \cite{wfo96} and treat hydrodynamic effects \cite{mtf03}.

A common simplifying assumption is that $\eta$, the ratio of trap depth
$\epsilon_t$ to sample temperature $k_B T$  is
large, where $k_B$ is the Boltzmann constant. For example, this yields analytic expressions for thermodynamic
quantities, and allows approximation of optical dipole traps \cite{gwo00} as
parabolic potentials \cite{cfs06}. By taking advantage of the high degree of
spatial symmetry in a linear potential, analytic expressions for thermodynamic
quantities were derived for the low-$\eta$ situation ($\eta<4$) in this
particular geometry \cite{ddo04}. It is worth emphasizing that Luiten's model
\cite{lrw96} is, in principle, valid for low $\eta$ as long as the assumptions
of ergodicity and a truncated Boltzmann distribution are also valid.

If the potential lacks the ideal shape of a power-law trap, simple analytic
expressions for many quantities of interest cannot be found, and numerical
methods are required. This is the case for low $\eta$  in an optical dipole
trap and especially when gravity is significant.  $^{88}$Sr in an optical
dipole trap falls into this situation because of its large mass and extremely
small $s$-wave scattering length $a_{88}=-1.4(6)\,a_0$ \cite{mmp08}, where the Bohr
radius $a_0 \approx 0.53$ \AA. Here, we describe numerical
methods appropriate for modeling collisional dynamics in an arbitrary trap in the low or high $\eta$ regime, which can be used for $^{88}$Sr.
Our approach builds on the works of  Luiten \textit{et al.} \cite{lrw96} and Comparat
\textit{et al.} \cite{cfs06}. As a check of the model, we also compare predictions with measurements of forced evaporation in  $^{84}$Sr, which has an
 $s$-wave scattering length of $a_{84}=122.7(3)\,a_0$ \cite{mmp08} and attains a much higher $\eta$, which allows direct evaporation to quantum degeneracy \cite{mmy09,sth09}. This model has also been used to interpret data on collisions involving Sr atoms in metastable states \cite{tcm08} and evaporative cooling of $^{87}$Sr and $^{88}$Sr for quantum degeneracy studies \cite{mmy10,dym10}. The main assumptions are ergodicity and the appropriateness of truncated Boltzmann distributions.

This paper is organized as follows. Section \ref{section:Experimental Setup}
describes the experimental setup, and then section \ref{section:Numerical
Simulation of Collisional Dynamics} presents the collisional processes
important in the trapped sample and the differential equations for evolution of
$N$ and $E$.
The numerical calculation is described in section \ref{Section:Description of Numerical Method}, and applications  of the model
to describe  trapped $^{88}$Sr and $^{84}$Sr, are discussed in section \ref{Numerical code and
simulation results}. The appendix describes an approximate treatment of the energy dependence of the $^{88}$Sr elastic collision cross section.


\section{Experimental Setup\label{section:Experimental Setup}}

 \begin{figure}
\includegraphics[keepaspectratio=true,width=3.3in,clip=true, trim=0 0 0 0, angle=0]{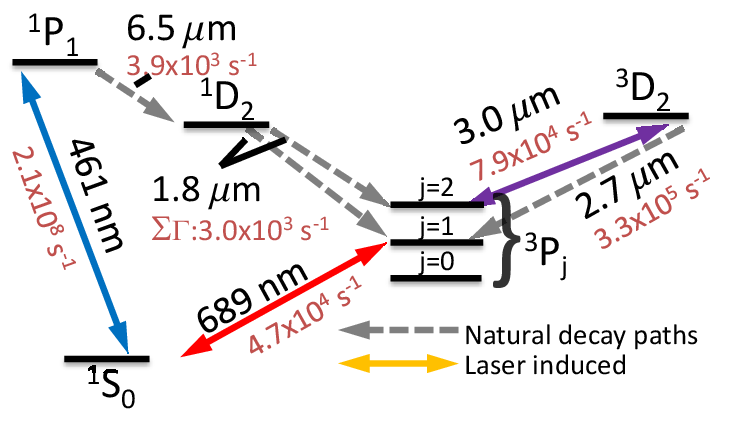}\\ 
  \caption{(Color online) Atomic Sr energy levels involved in laser-cooling.
  Decay rates (s$^{-1}$)
   and  excitation wavelengths are given for selected transitions.
  Laser light used for the experiment is indicated by solid lines.
  Atoms decaying to the $^3P_2$ level may be repumped by $3$\,$\mu$m light.
  }\label{figure:levels}
\end{figure}

The creation of samples of $^{88}$Sr or $^{84}$Sr atoms in an optical dipole trap (ODT) starts with laser cooling
and trapping phases that have been described in detail previously
\cite{nsl03,nms05,mms05,mmy09}.
Atoms are trapped in a magneto-optical trap (MOT) operating on the 461 nm
$^1S_0$-$^1P_1$ transition (Fig.\ \ref{figure:levels}) and cooled to about
2\,mK. There is a decay channel from the $^1P_1$ state, to the $^1D_2$ state
with a branching ratio of $2 \times 10^{-5}$. $^1D_2$ atoms can decay to the $^3P_1$ state, which decays to the ground state to allow further cooling, or to $^3P_2$ state, which can be trapped and accumulated in the magnetic trap formed by the quadrupole MOT magnets \cite{nsl03}.  $^3P_2$ atoms are repumped by applying a
3\,$\mu$m laser resonant with the ${^3P_2}$-${^3D_2}$ transition that returns
these atoms to the ground state \cite{mma09}. The repumped sample of atoms contains up to
$2.5 \times 10^8$ $^{88}$Sr atoms or $2.5 \times 10^7$ $^{84}$Sr atoms.

After this initial MOT stage, the 461 nm light is extinguished and
the atom sample is transferred with more than 50\% efficiency to a
second MOT operating on the $^1S_0$-$^3P_1$ intercombination line
\cite{kii99}. The atoms are cooled to 3\,$\mu$K in the $^{88}$Sr sample
or 1\,$\mu$K in the $^{84}$Sr sample, both producing peak
densities of $\sim 10^{12}$\,cm$^{-3}$.

\begin{figure}[htbp]
\centering
\includegraphics[trim = 0mm 0mm 0mm 0mm, clip, width=3.2in,height=2.25in]{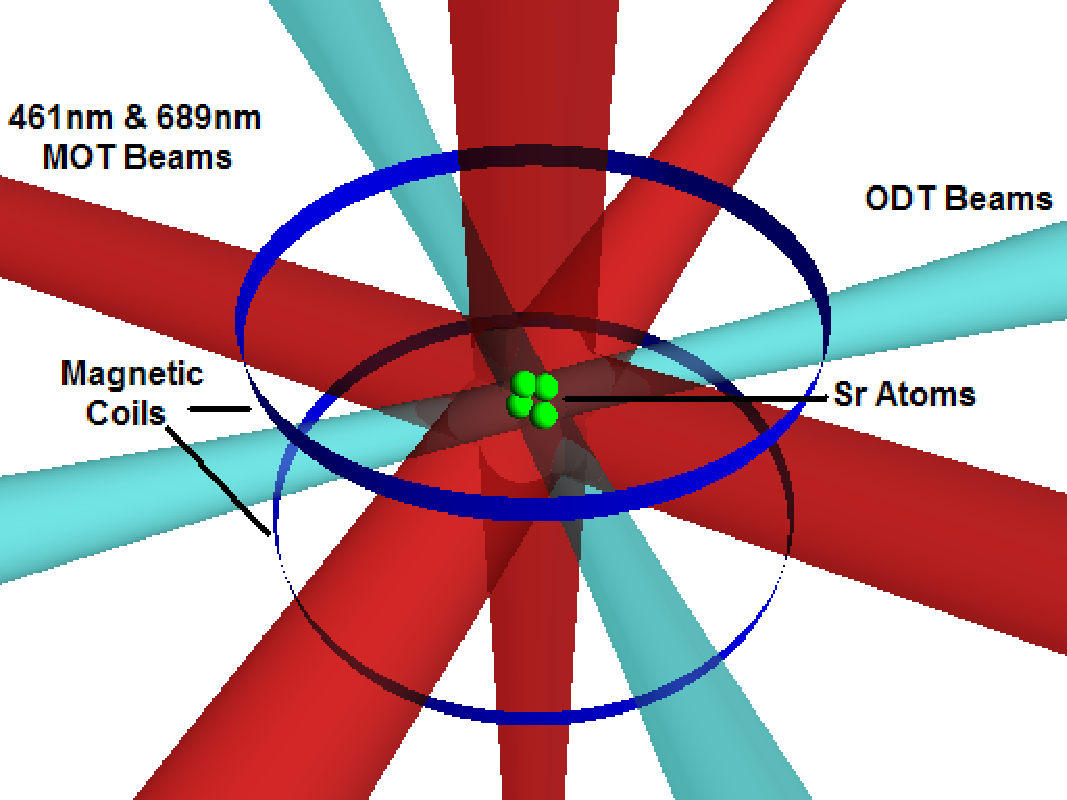}%
\caption{(Color Online) Schematic of our experiment illustrating the overlap of ODT beams
with MOT beams and relative positions of magnetic coils.}
\label{Figure: ODT Schematic}
\end{figure}

Atoms are then transferred to an ODT generated from a 21 W, 1064 nm, linearly-polarized,
  multi-longitudinal-mode fiber laser.
 The experimental setup is shown in Fig.\ \ref{Figure: ODT Schematic}.
 The trap is in a
crossed-beam configuration, derived from the first order deflection of an
acousto-optic modulator. The beam is focused on the atoms with a minimum
e$^{-2}$ intensity-radius of  $w\approx 100$\,$\mu$m. It is then reflected back
through the chamber to intersect the first beam at 90 degrees and refocused to
have approximately the same waist at the atoms. Both beams lie in a plane that
is inclined $10.5^{\circ}$ from horizontal.

The number of atoms and sample temperature  are determined with time-of-flight
absorption imaging using the $^1S_0$-$^1P_1$ transition. 
The ODT trapping potential is calculated from measured laser beam parameters
and the polarizability of the $^1S_0$ state \cite{ykk08}, and it is checked by
measuring the trap oscillation frequencies through the  parametric resonance
technique \cite{fdw98}. This allows us to infer the sample density profile from
the temperature and number of trapped atoms.

\section{Model of Collisional Dynamics
\label{section:Numerical Simulation of Collisional Dynamics}}

The evolution of
atom
number $N$ and total energy $E$ is described by a system of differential
equations. Different terms in the equations represent physical processes such
as elastic and inelastic collisions, and processes involving laser fields.

\subsection{Description of Basic Processes}

\subsubsection{Background collisions and inelastic collisional losses}
One-body losses due to collisions with background gas, and two- and three-body
inelastic collisional losses are described by the local equation
\begin{equation}\label{equation:local density evolution}
    \dot{n}_{coll}=-\Gamma_{bg} n-\beta_{in} n^{2}-Ln^{3},
\end{equation}
 where $n$ is the atomic
density. In simulations described here, we will assume that the loss rate
constants $\Gamma_{bg}$, $\beta_{in}$, and $L$
 are independent of
temperature. Integrating Eq.\ \ref{equation:local density evolution} over the
trap volume gives
\begin{eqnarray}\label{equation:number evolution in terms of N}
\dot{N}_{coll} & = & -\biggl(\Gamma_{bg}+\beta_{in}
N\frac{V_{2}}{V_{1}^{2}}+LN^{2}\frac{V_{3}}{V_{1}^{3}}\biggr)N,\end{eqnarray}
where  the effective volumes are
\begin{eqnarray}\label{equation:effective volumes}
V_{q} & \equiv & \frac{1}{n_{peak}^q}\int
d^{3}r[n(\mathbf{r})]^q,\end{eqnarray}
where $n_{peak}$ is the peak density in the trap.
We have also made use of the
relationship between peak density and total number, $n_{peak}V_{1}=N$.

The  energy or temperature evolution due to these processes for a constant trap
potential can be found as follows. The rate of energy change in an
infinitesimal volume $dV$ is
\begin{equation}\label{equation: local energy evolution}
d\dot{E}_{coll}=-\dot{n}_{coll}(\mathbf{r})dV[U(\mathbf{r})+\bar{E_{k}}(\mathbf{r})],
\end{equation}
where $U(\mathbf{r)}$ is the trap potential and $\bar{E_{k}}(\mathbf{r})$ is
the average kinetic energy per atom located at $\mathbf{r}$.  $U(\mathbf{r)}$
is defined to have a value of $U=0$ at the trap minimum. We assume a truncated
Boltzmann phase-space distribution (this is valid if the trap is sufficiently
ergodic \cite{lrw96}), which implies that the kinetic energy in a given
differential volume also obeys a truncated Boltzmann distribution truncated at
the kinetic energy required for an atom to escape the trap from the
differential volume. Thus the position-dependent average kinetic energy can be
expressed as
\begin{equation}\label{equation: local average kinetic energy}
 \bar{E_{k}}(\mathbf{r})=\frac{\int_{0}^{\epsilon_{t}-U(\mathbf{r})}dE_{k}
 E_{k}^{3/2}e^{-E_{k}/k_B T}}{\int_{0}^{\epsilon_{t}-U(\mathbf{r})}dE_{k}E_{k}^{1/2}e^{-E_{k}/k_B T}}.
\end{equation}

Integrating Eq.\ \ref{equation: local energy evolution} over the trap volume yields the rate of
change of total energy
\begin{eqnarray}
\dot{E}_{coll} & = &
-\Gamma_{bg}\biggl(\frac{T_{1}+P_{1}}{V_{1}}\biggr)N-\beta_{in}\biggl(\frac{T_{2}
+P_{2}}{V_{1}^{2}}\biggr)N^{2} \nonumber\\
&&-L\biggl(\frac{T_{3}+P_{3}}{V_{1}^{3}}\biggr)N^{3}.\end{eqnarray}  We have
introduced the effective kinetic energies
\begin{equation}\label{equation: effective kinetic energy}
  T_{q}\equiv\frac{1}{n_{peak}^{q}}\int d^{3}r[n(\mathbf{r})]^{q}\bar{E_{k}}
  (\mathbf{r}),
\end{equation}
and effective potential energies
\begin{equation}\label{equation: effective potential energy}
   P_{q}\equiv\frac{1}{n_{peak}^{q}}\int
d^{3}r[n(\mathbf{r})]^{q}U(\mathbf{r}).
\end{equation}
Note that the total energy of atoms inside the trap is
$E=(T_{1}+P_{1})\frac{N}{V_{1}}$.
To connect with a more intuitive expression, note that in the high-$\eta$ limit, $\bar{E_{k}}(\mathbf{r})=\frac{3}{2}k_B T$, which
can be taken out of the integrals to yield
\begin{eqnarray} \label{equation: energy loss for high eta}
\dot{E}_{coll}^{high-\eta} & \approx & -\frac{3}{2}k_B T\biggl(\Gamma_{bg}
N+\beta_{in}\frac{V_{2}}{V_{1}^{2}}N^{2}
+L\frac{V_{3}}{V_{1}^{3}}N^{3}\biggr)\nonumber \\
 &  & -\biggl(\Gamma_{bg}\frac{P_{1}}{V_{1}}N+\beta_{in}\frac{P_{2}}{V_{1}^{2}}N^{2}
 +L\frac{P_{3}}{V_{1}^{3}}N^{3}\biggr).
\end{eqnarray}

\subsubsection{Off-resonant laser scattering}

The scattering of  off-resonant photons, such as from the ODT laser, heats the
atoms due to momentum diffusion (MD) \cite{phi91}. The rate of change of total
energy due to this process is
\begin{eqnarray}
\dot{E}_{MD} & = & \Gamma_{laser}N E_{recoil},
\end{eqnarray}
where the total scattering rate of light from the laser field is given by
$\Gamma_{laser}$. If the light scattering rate is dominated by one transition,
$\Gamma_{laser}=\frac{s_0\gamma/2}{1+s_0+(2\delta_l/\gamma)^2}$, in which $s_0$
is the saturation parameter,  $\delta_l$ is the detuning of the laser, and
$\gamma$ is the linewidth of the transition. The recoil energy is
$E_{recoil}=k_B T_{recoil}=\hbar^2 k^2/m$, where $\hbar$ is
Planck's constant $h$ divided by $2\pi$, $k$ is the photon circular wavenumber,
and $m$ is the atom mass.

\subsubsection{Evaporation}

To describe the rate of atom loss due to evaporation, we follow the treatment
of \cite{lrw96}, which assumes ergodicity and a truncated Boltzmann
distribution in phase space,
\begin{eqnarray}\label{equation: truncated Boltzmann full distribution function}
f(\mathbf{r},\mathbf{p})&=&\frac{n_{0}}{(2\pi
mk_B T)^{3/2}}\exp\biggl[-\frac{U(\mathbf{r})+p^{2}/2m}{k_B T}\biggl]\nonumber
\\
&&\times \Theta(\epsilon_{t}-U(\mathbf{r})-p^{2}/2m).
\end{eqnarray}
$\Theta(\epsilon)$ is the Heaviside step function, and
$\mathbf{p}$ is the atom momentum with $p=\left|\mathbf{p}\right|$. Note that $n_{0}$  is not the peak density (the density at the trap minimum) unless the trap is infinitely deep. This yields a density distribution given by
\begin{eqnarray}\label{equation: density in terms of npeak}
n(\mathbf{r})&=&n_{peak}Ae^{-U(\mathbf{r})/k_B T}
\biggl\{\mathrm{erf}\biggl[\sqrt{\frac{\epsilon_{t}-U(\mathbf{r})}{k_B T}}\biggr]\nonumber
\\
&&-2\sqrt{\frac{\epsilon_{t}-U(\mathbf{r})}{\pi
k_B T}}\exp\biggl[-\frac{\epsilon_{t}-U(\mathbf{r})}{k_B T}\biggr]\biggr\},
\end{eqnarray}
 where the normalization constant $A$ is given by
\begin{equation}\label{equation: A}
A=\frac{n_0}{n_{peak}}=\biggl\{\mathrm{erf}\biggl[\sqrt{\frac{\epsilon_{t}}{k_B T}}\biggr]-2\sqrt{\frac{\epsilon_{t}}{\pi
k_B T}}\exp\biggl[-\frac{\epsilon_{t}}{k_B T}\biggr]\biggr\}^{-1}.
\end{equation}
The peak density is given by
\begin{eqnarray}\label{equation: npeak}
  n_{peak} &=& n(\mathbf{r})\vert_{U(\mathbf{r})=0} \nonumber \\
  &=& n_{0}\biggl\{\mathrm{erf}\biggl[\sqrt{\epsilon_{t}/k_B T}\biggr] \nonumber \\
  &&-2\sqrt{\epsilon_{t}/\pi
k_B T}\exp\biggr[-\epsilon_{t}/k_B T\biggr]\biggr\}.
\end{eqnarray}

The total number of atoms lost per unit time due to evaporation can then be
written as
\begin{equation}\label{equation: Ndot due to evap}
\dot{N}_{ev}=-\Gamma_{ev}N,
\end{equation}
where the evaporation rate per atom is
\begin{equation}\label{equation: evaporation rate constant}
\Gamma_{ev}=\frac{N}{V_{1}^{2}}A^{2}\sigma_{el}\bar{v}e^{-\eta}V_{ev}.
\end{equation}
Here, $\sigma_{el}$ is the elastic collision cross section, which is assumed to
be collision-energy independent in this treatment. $\bar{v}=\bigl(\frac{8k_B
T}{\pi m}\bigr)^{1/2}$ is the mean atomic velocity, and the effective volume
for elastic collisions leading to evaporation is
\begin{equation}\label{equation: effective volume for evaporation}
V_{ev}=\frac{\Lambda^{3}}{k_B T}
\int_{0}^{\epsilon_{t}}d\epsilon\rho(\epsilon)[(\epsilon_{t}-\epsilon-k_B T)e^{-\epsilon/k_B T}
+k_B Te^{-\eta}],
\end{equation}
where \begin{eqnarray}\label{equation: the thermal de Broglie wavelength}
\Lambda=(2\pi{\hbar}^2/mk_B T)^{1/2}
\end{eqnarray}
is the thermal de Broglie wavelength.
The density of states in the trap is given by
\begin{equation}\label{equation: density of states}
\rho(\epsilon)=\frac{2\pi(2m)^{3/2}}{h^{3}}\int_{U(\mathbf{r})\leq\epsilon_{t}}d^{3}
r\sqrt{\epsilon-U(\mathbf{r})}.
\end{equation}

Similarly, the rate of change of total energy due to evaporation is
\begin{equation}\label{equation: energy change due to evaporation}
\dot{E}_{ev}=-\Gamma_{ev}N\bar{E}_{ev},
\end{equation}
 where the average energy loss per
evaporated atom is
\begin{equation}\label{equation: energy loss per atom due to evap}
\bar{E}_{ev}=\epsilon_{t}+\frac{V_{ev}-X_{ev}}{V_{ev}}k_B T,
\end{equation}
with
\begin{equation}\label{equation: excess evaporation energy}
X_{ev}=\frac{\Lambda^{3}}{k_B T}\int_{0}^{\epsilon_{t}}d\epsilon\rho(\epsilon)[k_B Te^{-\epsilon/k_B T}
-(\epsilon_{t}-\epsilon+k_B T)e^{-\eta}].
\end{equation}

We note that the assumption of ergodicity that underlies this treatment  is equivalent to assuming three-dimensional evaporation, or that any atom with an energy greater than the trap depth escapes the trap before suffering a collision. This assumption is questionable when evaporation is over a saddle point, such as when gravity significantly modifies the potential. However, recent experiments in a similar trap geometry \cite{hzg08} to ours have shown that the evaporation efficiency can be near the three-dimensional limit if the trap is sufficiently asymmetric and non-separable, which is the case here. Hydrodynamic effects can also limit evaporation efficiency when the collisional mean free path is on the order of or smaller than the sample size \cite{cfs06}, but our experiments do not approach this regime, and we neglect these effects here.

\subsubsection{Time- dependent traps : forced evaporation}

When the trap confinement is varied adiabatically, such as during forced evaporation in an ODT when the trap-laser intensity is decreased, there is also an energy change due to reduction in the potential energy \cite{ber97,ogg01,cfs06}. This energy change can be expressed as
\begin{equation}\label{equation: energy change due to evaporation}
\dot{E}_{pot}=-\Gamma_{pot}N P_1/V_1,
\end{equation}
 where
$\Gamma_{pot}=\dot{U}/U$ and $P_1/V_1$ is the average potential energy per atom. In most experiments  with forced evaporative cooling, $\eta$ is relatively high, and $\Gamma_{pot}$ can be calculated using a harmonic approximation of the trap. For an isotropic trap, $U(r)=\frac{1}{2}m\omega^2r^2$ and
$\Gamma_{pot}=2\dot{\omega}/\omega$. For a nonisotropic potential,  $\omega$ is taken as the geometric mean of the angular oscillator frequencies \cite{cfs06}. When describing evaporation of $^{84}$Sr in Sec.\ \ref{Numerical code and simulation results}, we use this approximation for $\Gamma_{pot}$, with $P_1$, $V_1$, and $\omega$ found numerically for the trap as a function of ODT laser intensity.

\subsubsection{Final equations}

Accounting for all processes, the equations for number and energy evolution
become
\begin{eqnarray}\label{equation: Final equations} \dot{N} & = & -\Gamma_{bg}
N-\frac{1}{V_{1}^{2}}\bigl(\beta_{in}
V_{2}+A^{2}\sigma_{el}\bar{v}e^{-\eta}V_{ev}\bigr)N^{2}\nonumber \\
&&-L\frac{V_{3}}{V_{1}^{3}}N^{3},\\
\label{equation: Final equations 2}
\dot{E} & = & -\Gamma_{bg}\biggl(\frac{T_{1}+P_{1}}{V_{1}}\biggr)N
-\beta_{in}\biggl(\frac{T_{2}+P_{2}}{V_{1}^{2}}\biggr)N^{2} \nonumber \\
&&-L\biggl(\frac{T_{3}+P_{3}}{V_{1}^{3}}\biggr)N^{3} +\Gamma_{laser}N E_{recoil}\nonumber \\
 &  & -\frac{N^{2}}{V_{1}^{2}}A^{2}\sigma_{el}\bar{v}e^{-\eta}V_{ev}\biggl[\epsilon_{t}
 +\frac{V_{ev}-X_{ev}}{V_{ev}}k_B T\biggr]\nonumber \\
 && +\frac{2\dot{\omega}}{\omega}\frac{P_1N}{V_1}.
\end{eqnarray}

\section{Description of the Numerical Procedure}
\label{Section:Description of Numerical Method}
 Equations \ref{equation: Final equations} and \ref{equation:
Final equations 2} and the quantities contained therein provide a complete description of the evolution of the trapped gas within the approximations of ergodicity and a truncated Boltzmann distribution. Approximations are usually made to arrive at analytic results for required quantities (e.g. \cite{cfs06}) in order to facilitate solution of the differential equations. This is
straightforward for high-$\eta$ conditions \cite{kvd96}, and also in situations
of low-$\eta$ and with sufficient trap symmetry \cite{ddo04}.
For low-$\eta$ conditions and traps that lack spatial symmetry, numerical
evaluation of statistical mechanical quantities is the only option, and that is
the approach we follow, with the exception of calculation of $\Gamma_{pot}$. Numerical
evaluation is essential to describe our experiments with $^{88}$Sr in an ODT
because of the small scattering rate for this isotope, the importance of gravity, and the small inclination of our trap
lasers away from horizontal, which makes the trap very asymmetric. We perform all calculations in   $Mathematica^{TM}$.

\begin{figure}[htbp]
\centering
\includegraphics[trim = 15mm 0mm 15mm 25mm, clip, width=1.6in,height=1.4in]{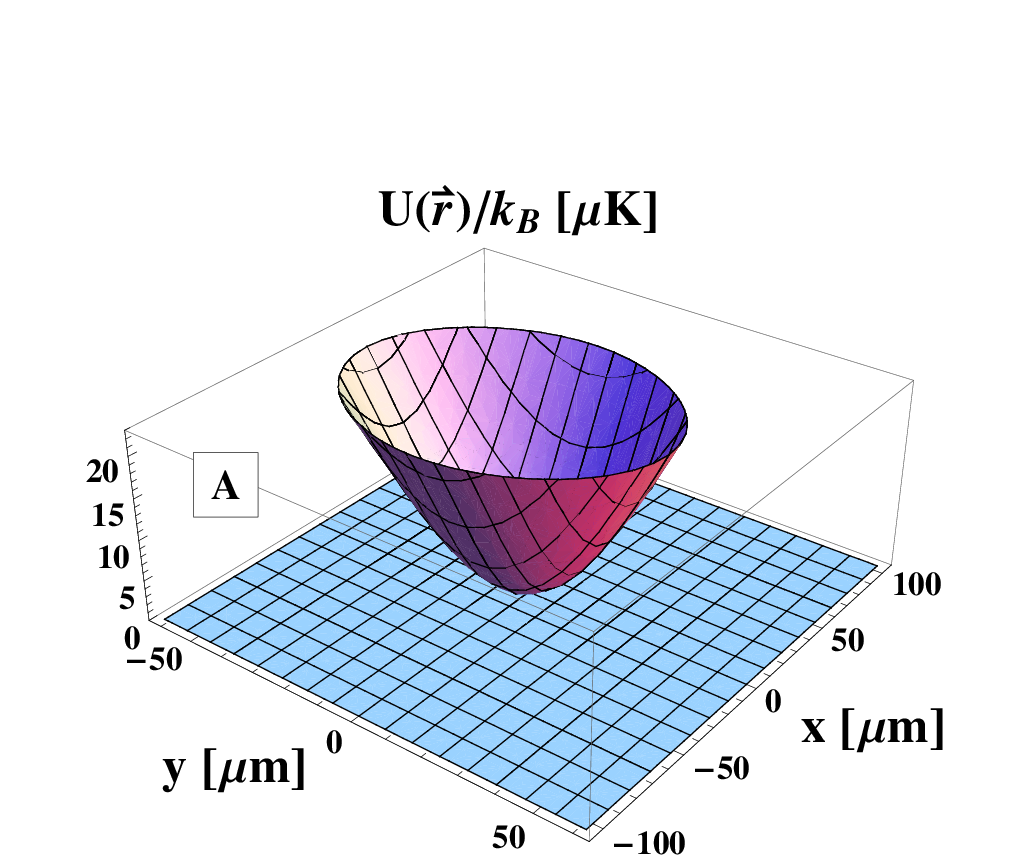}%
\includegraphics[trim = 15mm 0mm 15mm 25mm, clip, width=1.6in,height=1.4in]{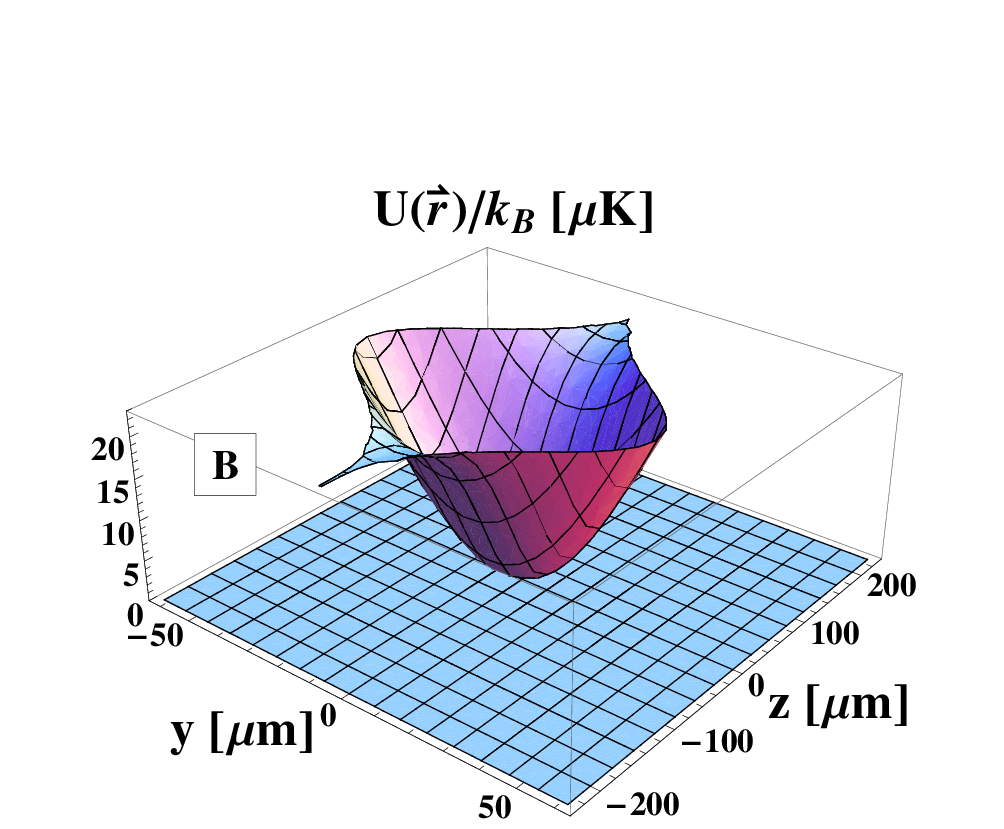}%
\caption{(Color online) (A) The x-y cut and (B) the y-z cut of the ODT potential for
approximately 9\,W per beam. The boundary of the trap is the set of points
with potential energy equal to the lowest saddle point, which is along the z
axis. Gravity is oriented in the -$\mathbf{\hat{y}}$ direction. The beam nearly
parallel to the x axis is slightly weaker and less focused than the beam along
z, leading to the observed asymmetry.
} \label{Figure:The potential for ODT}
\end{figure}

The first step in the  procedure is to find an
appropriate expression for the potential, $U(\textbf{r})$, at a given ODT laser intensity, for input to the numerical calculations.  Starting from the theoretical expression for the optical potential \cite{gwo00} created by the known atomic polarizability \cite{ykk08} and laser wavelength, powers, and waists, we employ an algorithm to find the trap minimum, the trap depth ($\epsilon_t$),
and the saddle points. This defines the trap boundaries and allows us to offset
the trap so that the minimum is $U=0$. An example is shown in Fig.\
\ref{Figure:The potential for ODT}.

 The formula for $U(\textbf{r})$ and description of the boundary is then passed to numerical integration
 routines for calculating statistical mechanical quantities described in section
\ref{section:Numerical Simulation of Collisional Dynamics} at a given temperature.  For  spatial integrals for  $V_q$ (Eq.\ \ref{equation:effective volumes}), $T_q$(Eq.\ \ref{equation: effective kinetic energy}),  $P_q$(Eq.\ \ref{equation: effective potential energy}), and $\rho(\epsilon)$ (Eq.\ \ref{equation: density of states}),
the integration extends over the entire region contained in the trap as determined with the algorithm described above. (This region is the connected region of space with $U(r)< \epsilon_t$ that contains the trap minimum.) An interpolating function representing $\rho(\epsilon)$ is used in evaluation of
$V_{ev}$ (Eq.\ \ref{equation: effective volume for evaporation}) and $X_{ev}$ (Eq.\ \ref{equation: excess evaporation energy}) in integrals over an energy interval from 0 to $\epsilon_t$.
An  adaptive-step-size integration routine in $Mathematica^{TM}$ is used to evaluate these integrals.

To provide a check of our programs, we compared the results of numerical calculations of all statistical mechanical quantities  for power-law traps to various analytic expressions that are available  in situations with such a simple form of the potential \cite{yan10}. Expressions in terms of incomplete Gamma functions can be obtained for power-law traps without making a high-$\eta$ approximation \cite{lrw96}. These are exact within  the truncated Boltzmann distribution approximation,  and our numerical results agree with them exactly. In the high-$\eta$ regime, simple analytic expressions can be found by making a high-$\eta$
approximation  \cite{kvd96,ddo04}. We find excellent agreement between our numerical calculations and these analytic expressions within their regime of validity.

The statistical mechanical quantities vary with  temperature and ODT laser intensity,
so integrals are evaluated at a dense series of temperature and laser intensity points.
The variation with temperature and laser intensity is
used to find  interpolating functions for the temperature and laser intensity dependence of all quantities,
which can then be used in place of time-intensive integral evaluations. Additionally, for
time-dependent traps, lookup
tables of trap depth and geometric average of the angular oscillator frequencies, which are the functions of ODT laser intensity, are necessary. It is important to note that the variation of $U$ and all quantities calculated from $U$ with ODT laser intensity allows us to model forced evaporation, since the ODT laser intensity is varied in a known way with time during the evaporation trajectory.

Using the interpolating functions, the atom number and temperature evolution can
easily be found for a given initial condition from Eqs.\ \ref{equation: Final equations} and \ref{equation:
Final equations 2} using an ordinary differential equations (ODE) solver in
$Mathematica^{TM}$. All terms are either constants or functions of number and temperature  and the independent variable time. This includes the total energy, $E$, so the ODE solver solves for $N(t)$ and $T(t)$ for the particular initial conditions and experimental parameters. Typically, one day is needed to create all lookup tables.
After this preparation, a complete ODE solution for
tens of seconds of sample evolution only requires a few seconds of computer
evaluation time. The programs used for these simulations  are available upon request.


\section{Evaporation of $^{88}$Sr and $^{84}$Sr} \label{Numerical code and simulation results}


\begin{figure}
  \includegraphics[keepaspectratio=true,width=3.5in,clip=true, trim=0 0 0 0, angle=0]{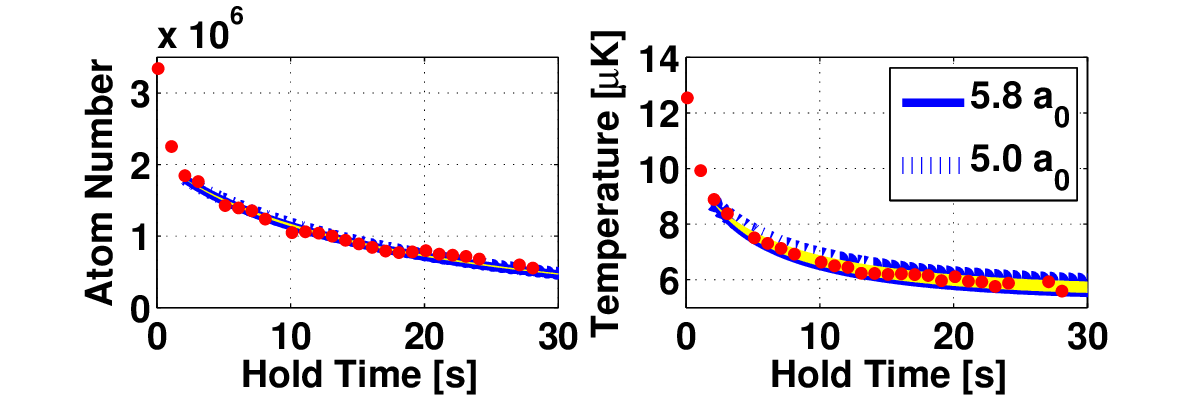}\\ 
  \caption{(Color online) Variation of atom temperature and number with time for $^{88}$Sr
  atoms in the $^1S_0$ state in a constant potential with a trap depth of 36\,$\mu$K. The solid curve shows the fitting result for the upper bound of
  the $s$-wave scattering length of $^{88}$Sr $a_{88}=5.8\,a_0$, while the dashed one shows that for the lower
  bound of $a_{88}=5.0\,a_0$.
  }\label{figure:Variation of 88Sr Atom Temperature and Number with Time}
\end{figure}


\begin{figure}
  \includegraphics[keepaspectratio=true,width=3.5in,clip=true, trim=0 0 0 0, angle=0]{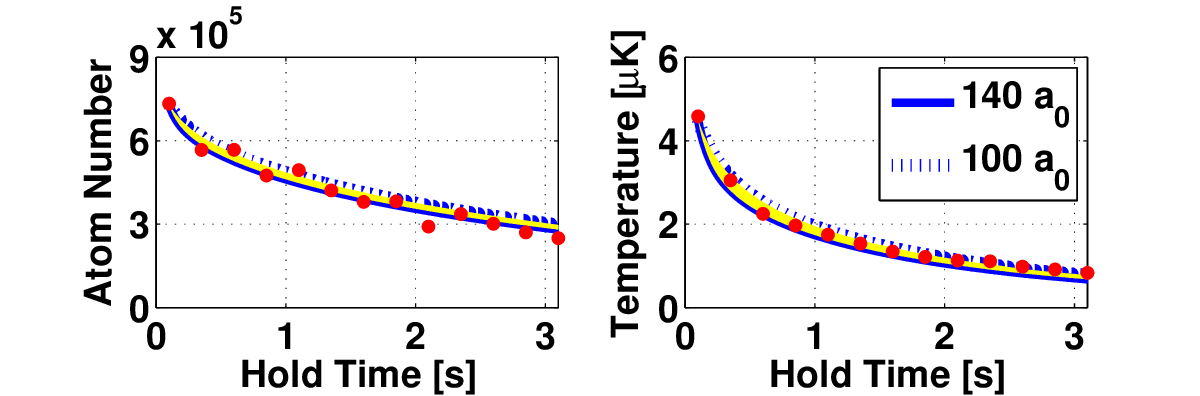}\\ 
  \caption{(Color online) Variation of atom temperature and number with time for $^{84}$Sr
  atoms in the $^1S_0$ state in a time-dependent trap. The solid curve shows the fitting result for the upper bound of
  the $s$-wave scattering length of $^{84}$Sr $a_{84}=140\,a_0$, while the dashed one shows that for the lower
  bound of $a_{84}=100\,a_0$.
  }\label{figure:Variation of 84Sr Atom Temperature and Number with Time in a time-dependent trap}
\end{figure}


Figure\ \ref{figure:Variation of 88Sr Atom Temperature and Number with Time} shows the
number and temperature of atoms as functions of time for $^{88}$Sr  $^1 S_0$
atoms in the ODT with a constant potential. Various quantities such as the one-body loss rate, two-body
inelastic collision rate constant, and elastic cross section can be
determined by fitting the calculated evolution curves to experimental data.
Three-body inelastic collisional loss is negligible here
due to the extremely small three-body loss-rate constant $L$ of $^{88}$Sr \cite{fdp06}.

 We exclude the first
second from the fit because we expect that atoms are far from equilibrium
and significant population is still trapped in the individual  beams of the ODT
and not in the crossed region at this time.
For this fit, we obtain the one-body loss rate
$\Gamma_1=0.04\,\mathrm{s}^{-1}$, the ODT photon scattering rate
$\Gamma_{ODT}=$ $0.03\,\mathrm{s}^{-1}$, and the $s$-wave scattering length of $^{88}$Sr
$a_{88}$ to an uncertainty of $\pm\,0.4\,a_0$ (Fig.\ \ref{figure:Variation of 88Sr Atom Temperature and Number with Time}), but uncertainty in trap waists of $\pm\,5\,\mu$m increases the uncertainty of $a_{88}$, and we quote a final value of $a_{88}=5.4\,^{+0.8}_{-0.6}\,a_0$ (the elastic scattering cross section $\sigma_{el}^{88}=730^{+230}_{-150}\,a_0^2$), which matches what is predicted by theory based on photoassociative spectroscopy data \cite{mmp08}. Note that at the temperature of the sample studied here, $\sigma_{el}$ differs significantly from its zero-temperature value, but sample temperature variation is small enough that the approximation of a constant cross-section during the simulation can describe the data well. (Appendix \ref{section:88Sr Elastic Cross Section} describes how the energy dependence of the cross section is accounted for when comparing $\sigma_{el}$ determined from this analysis with theory.)
A good fit is found with $\beta_{in}=0$
as expected since there are essentially no inelastic two-body loss processes in this
system.


Figure\ \ref{figure:Variation of 84Sr Atom Temperature and Number with Time in a time-dependent trap} shows the simulation
and data for $^{84}$Sr $^1S_0$ atoms in a time-dependent trap.
The power of ODT beams is ramped down according to $P=P_0/(1+t/\tau)^\beta$, with time denoted by $t$, $\beta=1.5$, and $\tau=$ $2\,s$, and the trap depth is reduced from 36\,$\mu$K initially to 5\,$\mu$K within 3.1\,s. The peak phase space density during this interval is about 0.06, so effects of quantum degeneracy can be neglected.

The $s$-wave scattering length of $^{84}$Sr is $a_{84}=122.7(3)\,a_0$ \cite{mmp08}, so evaporation is much more efficient than for $^{88}$Sr. Due to the large scattering length, three-body loss also becomes important. In this fit, we assume a value of the three-body loss-rate constant, $L^{84}=3 \times 10^{-27}$ cm$^6$/s, which is found from the measured value for $^{86}$Sr \cite{stg10} and the $a^4$-scattering-length dependance of $L$ \cite{egb99}.
The fit determines $a_{84}$ to an uncertainty of $\pm\,20\,a_0$ (Fig.\ \ref{figure:Variation of 84Sr Atom Temperature and Number with Time in a time-dependent trap}), but uncertainty in trap waists of $\pm\,5\,\mu$m increases the uncertainty of $a_{84}$, and we quote a final value of $a_{84}=120\,^{+30}_{-40}\,a_0$, which is in good agreement with previous determinations \cite{mmp08,skt10}.
We are relatively insensitive to the values of $\Gamma_1$ and $\Gamma_{ODT}$ because the sample evolution is fast, so we set these parameters to values implied by $^{88}$Sr data.
The agreement with the experimental value of $a_{84}$ confirms the validity of the model for time-dependent traps, which was valuable for guiding recent experiments  attaining  quantum degeneracy in Sr \cite{mmy10,mmy09}.

\section{Conclusion\label{section:Conclusion}}
In this paper, we have presented a model describing inelastic and elastic
collision dynamics of trapped atoms that can treat traps lacking spatial
symmetry and samples with a wide range of $\eta$, especially in low-$\eta$ conditions.
The main assumptions are ergodicity and a
truncated Boltzmann velocity distribution.
The
model was used to describe $^{88}$Sr and $^{84}$Sr in an asymmetric ODT with low
$\eta$ and high $\eta$ respectively, and collisional parameters extracted from the data were found to agree
well with those from previous works. This model has been used to
extract elastic and inelastic cross sections from experiments  with metastable Sr atoms in an optical dipole trap \cite{tcm08} and
to guide  achievement of quantum degeneracy \cite{mmy09,mmy10}.

\appendix

\section{Energy Dependence of the $^{88}$Sr Elastic Collision Cross Section \label{section:88Sr Elastic Cross Section}}

\begin{figure}
\includegraphics[width=3.4in,clip=true, trim=0 0 0 0, angle=0]{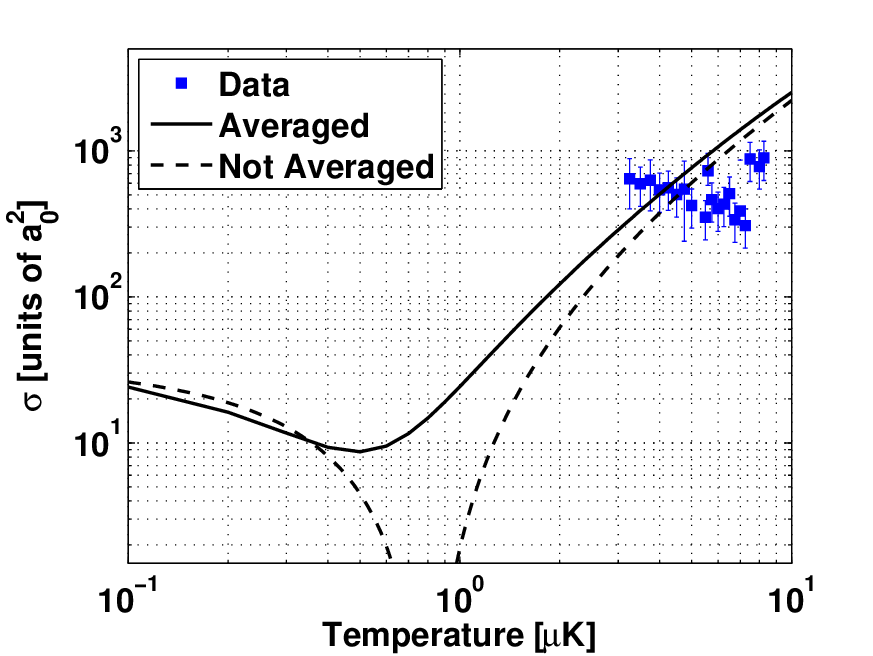}\\
  \caption{Measurements and theories for the $^{88}$Sr elastic collision
  cross section, $\sigma_{el}$. The dashed curve is the energy-dependent elastic collision cross section \cite{mmp08}
  for a collision energy of $E_{coll}=2k_B T$, which is an average collision
  energy as described in the text. The solid line is an average cross section,
  weighted by the rate of collisions of a given energy,
  for the temperature $T$.
  }\label{figure:Sr88 scattering length}
\end{figure}

Evaporation with an energy-dependent elastic collision cross section has been
modeled in \cite{thv04,cfs06}, but these treatments assume a large cross
section that varies because of the unitarity limit.  In $^{88}$Sr, the cross
section varies because the scattering length is very small, as shown in
Fig.\ \ref{figure:Sr88 scattering length} \cite{mmp08}. This variation is significant
 at microkelvin energies, which complicates comparison of
theory and experiment because a distribution of collision energies contributes
in a thermal sample in the ODT. A full energy-dependent kinetic calculation is
beyond the scope of this model. We treat the variation in approximate fashion
by assuming the system can be described by an effective, temperature-dependent
cross section, $\langle \sigma_{el}\rangle$, that  is an average of the
collision-energy dependent cross section.

To relate $\langle \sigma_{el}\rangle$ to the underlying energy-dependent
cross section, first consider the number of elastic collisions per second per unit
volume at position $\bf{r}$ \cite{huang,mcq}
\begin{equation}\label{equation: umber of elastic collsions per second per volume}
 Z(\mathbf{r})=\frac{1}{2}\int d^{3}p_{1}\int
d^{3}p_{2}\sigma_{el}\vert\mathbf{v_{1}}-\mathbf{v_{2}}\vert
f(\mathbf{r},\mathbf{p_{1}},t)f(\mathbf{r},\mathbf{p_{2}},t),
\end{equation}
where $\mathbf{v_{1}}=\mathbf{p_{1}}/m$, $\mathbf{v_{2}}=\mathbf{p_{2}}/m$. For the ultracold regime,  $\sigma_{el}(E_{coll})$ can only depend on the collision energy $E_{coll}=p^2/2\mu$ for
  $p=\mu \left\vert \mathbf{v}_{1}-\mathbf{v}_{2} \right\vert$ and the reduced mass $\mu=m/2$.
 The average cross-section will not depend on density, so we can assume a constant density.
For simplicity, we use untruncated Maxwell-Boltzmann distributions for $f$, which yields
\begin{eqnarray}\label{equation:Z}
Z &=& \frac{2\pi n_{0}^{2}}{\mu}\int dp\, p^3 \frac{\sigma_{el}
(E_{coll})}{(2\pi \mu k_B T)^{3/2}} e^{-\frac{p^2}{2\mu k_B T}}.
  \end{eqnarray}
For an energy-independent cross section, this would reduce to the standard result \cite{huang}
\begin{eqnarray}\label{equation:HuangZII}
    Z=2\sigma_{el} n_{0}^{2}\sqrt{\frac{k_B T}{\pi m}}.
\end{eqnarray}

For a given sample temperature, there is a distribution of collision energies, and if the cross section is energy dependent, this would imply a distribution of cross sections. We treat this possibility, which is important for $^{88}$Sr collisions, by calculating an effective
cross section, $\langle \sigma_{el}\rangle$, that can be used in our simulations. This quantity is an average of the
true cross section over energy, and in principle it can vary with temperature.

To arrive at a value for $\langle \sigma_{el}\rangle$ from the theoretical $\sigma_{el}(E_{coll})$
for $^{88}$Sr \cite{mmp08} we assume that the evaporation rate can be described by an average of $\sigma_{el}(E_{coll})$
in which the weighting is proportional to the contribution of each collision energy to
the total number of collisions per time in the sample. (Using this weighting,
the average collision energy for a given temperature $T$ for an energy-independent cross section is $2k_B T$.) For a
given equilibrium temperature, this average cross section is given by
\begin{equation}\label{average cross section}
\langle \sigma_{el} \rangle =\frac{Z}{2n_0^2\sqrt{\frac{k_B T}{\pi m}}},
\end{equation}
where $Z$ is calculated numerically using Eq.\ \ref{equation:Z} and the  energy
dependence of $\sigma_{el}$ that was determined from photoassociation data \cite{mmp08}.  Contribution to $Z$ is not exactly
equivalent to contribution to the evaporation rate, but this is a reasonable
approximation in the spirit of \cite{thv04}.

Figure\ \ref{figure:Sr88 scattering length} shows variation of
$\sigma_{el}(E_{coll})$ and $\langle\sigma_{el}\rangle$ for $^{88}$Sr as well
as experimental data in which
the numerical model  is used to determine the best-fit $\sigma_{el}$ in Eq.\ \ref{equation: Final equations 2}. The reasonable match of theory and experiment
gives confidence in the numerical model in the low-$\eta$ regime. Error bars represent statistical
variation. In addition, there is systematic uncertainty due to uncertainty in
the trapping potential of typically a factor of two, but this becomes more of
an issue for lower sample temperatures and shallower traps, which are more
challenging to characterize and model due to the importance of gravity. The
assumption of ergodicity may also be less valid at lower temperature because
the elastic collision rate and $\eta$ become very small.

\section*{ACKNOWLEDGMENTS}

We thank P. Julienne, R. C\^{o}t\'{e}, and P. Pelligrini for helpful
discussions and D. Comparat for sharing the original code for describing
collisional dynamics of trapped atoms at high $\eta$. This work was supported
by the Welch Foundation (C-1579), National Science Foundation (PHY-0855642),
and the Keck Foundation.


\end{document}